\documentclass[aps, prd, reprint, nofootinbib, amsmath, amssymb, floatfix]{revtex4-2}

\usepackage{graphicx}
\usepackage[dvipsnames]{xcolor}
\usepackage{hyperref}
\hypersetup{colorlinks, urlcolor=BlueViolet, citecolor=Plum, linkcolor=PineGreen}

\usepackage{cleveref}
\usepackage[compat=1.1.0]{tikz-feynman}
\usepackage{slashed} 

\newcommand{\vev}[1]{\left< #1 \right>}      % vacuum expectation value

\begin{document}

\title{Absorption of Fermionic Dark Matter via the Scalar Portal}

\author{Peter Cox}\email{peter.cox@unimelb.edu.au}
\author{Matthew J. Dolan}\email{matthew.dolan@unimelb.edu.au}
\author{Joshua Wood}\email{jwood5@student.unimelb.edu.au}

\affiliation{ARC Centre of Excellence for Dark Matter Particle Physics, \\
School of Physics, The University of Melbourne, Victoria 3010, Australia}

\begin{abstract}
The absorption of fermionic dark matter has recently been studied as a signature for the direct detection of dark matter. We construct the first UV completion of the scalar effective operator associated with this signature. We calculate the constraints on the model and demonstrate there is viable parameter space which can be probed by a next-generation experiment such as XLZD. We also consider the cosmological history of our model and show that the correct relic abundance can be obtained via freeze-out in the dark sector. However, within this minimal model, we find that the absorption signal is highly suppressed in the parameter space that yields the correct relic abundance. 
\end{abstract}

\maketitle

%%%%%%%%%%%%%%%%%%%%%%%%%%%%%%%%%%%%%%%%%%%%%%%%%%%%

\section{\label{sec:intro}Introduction}

The direct detection of dark matter in terrestrial experiments remains one of the most promising strategies for understanding its microscopic properties. The traditional approach to direct detection involves searching for the elastic scattering of dark matter on nuclei $(\chi N \to \chi N)$, and has produced strong bounds on dark matter--nucleon interactions for dark matter masses in the GeV\,--\,TeV range~\cite{LUX-ZEPLIN:2022xrq}. However, within the landscape of dark matter models there are a range of potential direct detection signals beyond elastic nuclear (or electron) scattering. An interesting class of signals arise from inelastic processes in which the dark matter is absorbed.

Dark matter absorption processes are, of course, used extensively to search for light bosonic dark matter, for example axions and dark photons. On the other hand, the direct detection of fermionic dark matter through absorption has been comparatively less explored. This is despite the fact that it occurs generically in models where there is no stabilising symmetry for the dark matter. The requirement that dark matter is long-lived on cosmological timescales leads to strongly suppressed interactions unless the dark matter is sufficiently light. Hence, fermionic dark matter can yield an observable absorption signal only for keV-scale masses, once the lower bound of $m_\chi\gtrsim$~keV~\cite{Alvey:2020xsk,Carena:2021bqm} from Pauli exclusion is also taken into account. While keV-scale dark matter is challenging to detect in elastic scattering experiments, absorption processes have the advantage that the entire rest mass energy of the dark matter can be deposited in the detector. 

In this paper, we focus on fermionic dark matter absorption via interaction with electrons\footnote{Fermionic dark matter absorption via interaction with nucleons has been considered in \cite{Ando:2010ye,Dror:2019onn,Dror:2019dib}.}, specifically the process $e\chi \to e \nu$. This process was first considered in the context of sterile neutrino dark matter via active-sterile mixing~\cite{Ando:2010ye}; however, in that scenario the potentially accessible parameter space is excluded by X-ray constraints on decaying dark matter~\cite{Campos:2016gjh}. More recently, it was revisited in Refs.~\cite{Dror:2020czw,Ge:2022ius}, which used an effective field theory (EFT) approach to investigate the sensitivity of existing and future experiments. The PandaX-4T experiment has subsequently performed a dedicated search for this signal~\cite{PandaX:2022ood}.

An advantage of the EFT approach is its model-independence; however, one must also consider whether or not there exist viable ultraviolet (UV) completions. There are several additional, non-trivial requirements that the UV completion must satisfy: cosmological stability of dark matter and observational bounds on its decay; constraints on the mediator from both terrestrial experiments and astrophysics; a viable dark matter production mechanism and thermal history; and invariance under the full Standard Model (SM) gauge group. As we shall demonstrate, the combination of these considerations can be highly constraining, and may exclude the possibility of observing a signal in future direct detection experiments.

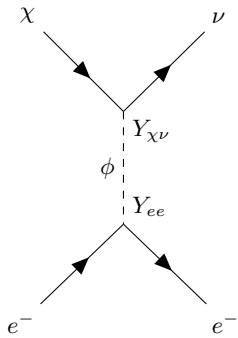
\begin{figure}
\begin{tikzpicture}
  \begin{feynman}
    \vertex [label=325:\(Y_{\chi\nu}\)](a);
    \vertex [above left=of a] (i1) {\(\chi\)};
    \vertex [above right=of a] (f1) {\(\nu\)};
    \vertex [below =of a, label=4:\(Y_{ee}\)] (b);
    \vertex [below left=of b] (i2) {\(e^-\)};
    \vertex [below right=of b] (f2) {\(e^-\)};

    \diagram* {
      (i1) -- [opacity=0.0] (f1),
      (i1) -- [fermion] (a) -- [fermion] (f1),
      (a) -- [scalar, edge label'=\(\phi\)] (b),
      (i2) -- [fermion] (b) -- [fermion] (f2),
      };
  \end{feynman}
\end{tikzpicture}
\caption{Dark matter absorption via electron interaction.}
    \label{fig:absorption-diagram}
\end{figure}

In this work, we present a minimal UV completion of the scalar\footnote{For a UV completion of the vector operator $(\bar{\chi}\gamma_\mu P_R\nu) (\bar{e}\gamma^\mu e)$ see~\cite{Dror:2020czw}. That paper also proposed a scalar-mediated simplified model, but this requires a UV completion to be invariant under the SM gauge group.} operator $(\bar\chi P_R \nu)(\bar e e)$. In addition to the Dirac fermion dark matter $\chi$, the model features a right-handed sterile neutrino and a scalar mediator that mixes with the Higgs, leading to the absorption process shown in \cref{fig:absorption-diagram}. The sterile neutrino is simultaneously responsible for generating Dirac neutrino mass.

There are several complementary probes of our model, some of which already provide strong bounds on the parameter space. First, the decay $\chi\to\nu\gamma\gamma$ is tightly constrained by X-ray observations for dark matter masses above 15\,keV. There are also strong bounds on the scalar mediator from horizontal branch stars and supernova 1987A. Taking these constraints into account, we show that there remains parameter space that could be probed by a next-generation liquid xenon direct detection experiment, if one ignores the additional requirement of a viable dark matter production mechanism. 

In much of the parameter space, including the region of potential interest for direct detection, the dark sector self-thermalises and the relic abundance is produced by freeze-out in the dark sector. We show that this mechanism can yield the observed relic abundance, but simultaneously predicts a highly suppressed direct detection signal. Contrary to the na\"ive expectations of the EFT analysis, we conclude that in a minimal UV completion direct detection is not a powerful probe of the model.

The structure of the paper is as follows. In \cref{sec:model} we present our UV completion of the scalar operator. \Cref{sec:phenomenology} discusses the phenomenology of the model, including dark matter decays, and collider and astrophysical bounds on the scalar mediator. Production of the dark matter is addressed in \cref{sec:production}, before we conclude in \cref{sec:conclusion}.

%%%%%%%%%%%%%%%%%%%%%%%%%%%%%%%%%%%%%%%%%%%%%%%%%%%%

\section{\label{sec:model}The Model}

We begin by introducing a UV complete model that can, in principle, yield an observable direct detection signal. At the level of the EFT, the scalar-mediated dark matter absorption process $e\chi \to e \nu$ proceeds via the dimension-6 operator/s
\begin{equation}
    \frac{1}{\Lambda^2}(\bar{\chi}P_{L,R}\nu)(\bar{e}e) \,,
\end{equation}
where $\Lambda$ is a scale associated with the UV completion of the effective operator. The tree-level completion of this operator is shown in \cref{fig:absorption-diagram}, with $\Lambda$ given in terms of UV parameters by
\begin{equation}
    \frac{1}{\Lambda^4} = \frac{\left|Y_{ee}Y_{\chi \nu}\right|^2}{m_\phi^4} \,,
\end{equation}
where $Y_{ee}$ and $Y_{\chi\nu}$ are the vertex factors depicted in the diagram.

The minimal UV completion features the SM Higgs as the scalar mediator and introduces a coupling to dark matter of the form $y_{H\chi} \bar{L} \Tilde{H} \chi$. However, this scenario cannot yield an observable signal in direct detection. First, assuming that $y_{H\chi}\sim\mathcal{O}(1)$, the effective operator scale is set by the electroweak scale, combined with the electron Yukawa coupling: $\Lambda\sim m_h/\sqrt{y_e}$. This is already beyond the foreseeable reach of direct detection experiments~\cite{Dror:2020czw}. Second, after electroweak symmetry breaking, the coupling $y_{H\chi}$ leads to mixing between the dark matter and the active neutrinos. This enables the radiative decay $\chi\to\nu\gamma$ to proceed via a $W$ boson/charged lepton loop. This dark matter decay mode is highly constrained by X-ray line searches, translating into an upper bound on the coupling of $y_{H\chi} \lesssim 10^{-26}$ for keV scale dark matter~\cite{Ng:2019gch}.

With the minimal scenario ruled out, we consider a new light scalar mediator. We introduce a real gauge singlet scalar field $\phi$ that couples to the SM fermions via mixing with the Higgs. We also introduce a chiral, gauge singlet fermion $\nu_R$ that couples to the dark matter via a term $\phi\bar{\nu}_R\chi$. As its name suggests, $\nu_R$ is also responsible for the generation of neutrino mass. After integrating out $\phi$, this model generates the operator $(\bar{\chi}P_R\nu)(\bar{e}e)$.

The Lagrangian describing the model is
\begin{equation}\label{eq:FullLagrangian}
    \mathcal{L} = \mathcal{L}_{SM} + \mathcal{L}_{\chi} + \mathcal{L}_{\nu_R} + \mathcal{L}_{\phi} + \mathcal{L}_{int} \,,
\end{equation}
with
\begin{align}
    \mathcal{L}_\chi & = i\bar{\chi}'_L\slashed{\partial}\chi'_L + i\bar{\chi}'_R\slashed{\partial}\chi'_R - M_\chi (\bar{\chi}'_L\chi'_R + h.c.) \,, \\
    \mathcal{L}_{\nu_R} & = i\bar{\nu}'_R\slashed{\partial}\nu'_R - \big(y_{H\nu}\bar{L}\Tilde{H}\nu'_R + h.c.\big) \,, \\
    \mathcal{L}_{\phi} & = \frac{1}{2}\partial_\mu \phi' \partial^\mu \phi' - \kappa_{1} \phi' - \frac{\kappa_2}{2}\phi'^2 - \frac{\kappa_3}{3!}\phi'^3 - \frac{\kappa_4}{4!}\phi'^4 \,, \\
    \begin{split}
    \mathcal{L}_{int} & = - \delta_1 H^\dag H \phi' - \frac{\delta_2}{2}H^\dag H \phi'^2 \\
    & - \left(y_{\phi\chi\nu} \phi'\bar{\chi}'_L \nu'_R + h.c.\right)
    - \left(M_{\chi\nu}\bar{\chi}'_L \nu'_R + h.c.\right) \,,
    \end{split}\label{eq:IntLagrangian}
\end{align}
where we use $^\prime$ to denote fields prior to diagonalisation to the mass basis, and $\Tilde{H} = i\sigma^2 H^*$. All couplings and masses are taken to be real without loss of generality.

The above Lagrangian exhibits a $U(1)$ lepton number symmetry, with the charges of the non-SM fields given by $L(\nu'_R) = L(\chi'_{L,R}) = + 1$. This has been imposed to forbid a Majorana mass term for $\nu_R$. For Majorana neutrinos, the dark matter absorption process in \cref{fig:absorption-diagram} would be suppressed by a small active-sterile mixing angle, since the outgoing neutrino state is required to be approximately massless. This symmetry also enforces $\chi'$ to be a Dirac fermion.

Another important feature of the Lagrangian is the absence of a $\bar{L} \Tilde{H} \chi'$ coupling, which is enforced by a $\mathbb{Z}_2$ symmetry under which $\chi'$ and $\phi'$ are odd, while all other fields are even. This $\mathbb{Z}_2$ is softly broken by $\kappa_1$, $\kappa_3$, $\delta_1$, and $M_{\chi\nu}$, allowing $\phi'$ to mix with the Higgs and for the $(\bar{\chi}P_R\nu)(\bar{e}e)$ operator to be generated\footnote{This can also be achieved through only spontaneous breaking of the  $\mathbb{Z}_2$ symmetry by $\langle\phi'\rangle$; this is a special case of the more general model.}, while still providing a natural suppression of the decay $\chi\to\nu\gamma$ (as discussed in \Cref{sec:phenomenology}).

Finally, for simplicity, we consider only a single neutrino generation, which for concreteness we take to be the electron neutrino. The generalisation to three generations is relatively straightforward and is not expected to significantly affect the phenomenology of the model, but would enhance the direct detection signal by a factor of three.

%%%%%%%%%%%%%%%%%%%%%%%%%%%%%%

\subsection{Diagonalisation and Mixing}\label{sec:Diagonalisation}
    
After electroweak symmetry breaking, we expand $H$ around its vacuum expectation value (VEV) using the usual parameterisation
\begin{equation}\label{eq:HVEV}
    H \equiv \frac{1}{\sqrt{2}} \begin{pmatrix}0 \\ v + h'\end{pmatrix} \,,
\end{equation}
where $v \approx 246$\,GeV and $h'$ is a real scalar field. We also re-express $\phi'$ as an expansion around its VEV (which may be zero): $\phi' \to \phi' + \vev{\phi'}$. Explicit expressions for the scalar VEVs and the resulting mass matrices in terms of the Lagrangian parameters are given in \cref{app:model_details}.

If either $\delta_1$ or $\delta_2 \left<\phi'\right>$ are non-zero, mixing occurs between the SM Higgs boson $h'$ and $\phi'$. The mixing can be expressed in terms of an angle $\theta_\phi$, such that
\begin{equation}
    \begin{pmatrix}h \\ \phi\end{pmatrix} =
    \begin{pmatrix}\cos\theta_\phi & -\sin\theta_\phi \\ \sin\theta_\phi & \cos\theta_\phi\end{pmatrix}
    \begin{pmatrix}h' \\ \phi'\end{pmatrix} \,,
\end{equation}
where $h$ and $\phi$ (unprimed) denote the heavier and lighter mass eigenstates respectively. We take $m_h \approx 125$\,GeV and $\phi$ is the light mediator relevant for dark matter absorption.

Mixing occurs between the dark matter and neutrino via the $M_{\chi\nu}$ and $y_{\phi\chi\nu} \left<\phi'\right>$ terms in the Lagrangian. The diagonalisation of the corresponding mass matrix is detailed in \cref{app:model_details}. The mass eigenstates (again unprimed) can be expressed in terms of real rotations:
\begin{align}
    \begin{pmatrix}
        \nu_{L,R} \\ \chi_{L,R}
    \end{pmatrix}
    & =
    \begin{pmatrix}
        \cos\theta_{L,R} & -\sin\theta_{L,R} \\
        \sin\theta_{L,R} & \cos\theta_{L,R}
    \end{pmatrix}
    \begin{pmatrix}
        \nu'_{L,R} \\ \chi'_{L,R}
    \end{pmatrix} \,.
\end{align}

In the massless neutrino limit, $\sin\theta_L$ vanishes and mixing only occurs between $\chi_R'$ and $\nu_R'$. After diagonalisation, and working in this limit, the tree-level couplings between $\chi_{L,R}$ and $\nu_{L,R}$ are given by
\begin{align}
    \mathcal{L} \supset &- \cos\theta_R \cos\theta_\phi y_{\phi\chi\nu}(\phi \bar{\chi}_L\nu_R) + h.c.\nonumber\\
    &- \sin\theta_R \cos\theta_\phi y_{\phi\chi\nu}(\phi \bar{\chi}_L\chi_R) + h.c.
\end{align}
The vertex factors contributing to the dark matter absorption process in \cref{fig:absorption-diagram} can then be written explicitly:
\begin{align}
    Y_{ee} & = \frac{m_e}{v}\sin\theta_\phi \,,\\
    Y_{\chi\nu} & = y_{\phi\chi\nu}\cos\theta_R \cos\theta_\phi \,.
\end{align}
See \cref{app:model_details} for the full expressions, including additional couplings proportional to neutrino mass.

%%%%%%%%%%%%%%%%%%%%%%%%%%%%%%%%%%%%%%%%%%%%%%%%%%%%

\section{\label{sec:phenomenology}Phenomenology}

\subsection{\label{sec:direct_detection}Direct Detection}

In a direct detection experiment, the absorption of $\chi$ can transfer sufficient energy to ionise an atomic electron. This signal can be detected in large volume argon and xenon time projection chambers via subsequent scintillation and secondary ionisation. The differential rate for dark matter absorption via the operator $(\bar{\chi}P_R\nu)(\bar{e}e)/\Lambda^2$ was calculated in Refs.~\cite{Dror:2020czw,Ge:2022ius} for a xenon target. The event rate is proportional to the effective absorption cross-section,
\begin{equation}
    \sigma_e \equiv \frac{m_\chi^2}{4\pi\Lambda^4} \,.
\end{equation}
The projected sensitivities of current and proposed future xenon experiments to $\sigma_e$ were calculated by Ref.~\cite{Dror:2020czw}. The authors assumed a detection efficiency equal to that of the XENON1T experiment~\cite{XENON:2020rca}, with an energy threshold of 1\,keV (sensitive to absorption events for $m_\chi \gtrsim 3$\,keV), and determined the projected sensitivity by requiring at least 10 events during an experiment's full exposure. Here, we predominantly focus on the sensitivity at the proposed XLZD detector~\cite{DARWIN:2016hyl} (assuming an exposure of 200 tonne-years), which offers the best prospects in the foreseeable future.

The typical momentum transfer in the absorption process is of order $m_\chi$. The EFT description is therefore valid provided $m_\phi \gg m_\chi$. In terms of the parameters of our model, the effective absorption cross-section in this limit is
\begin{align}
    \sigma_e & = \frac{m_\chi^2 Y_{ee}^2 Y_{\chi\nu}^2}{4\pi m_\phi^4}\\
    & = \sin^2\theta_\phi\cos^2\theta_\phi \cos^2\theta_R \left(\frac{m_e}{v}\right)^2 \frac{y_{\phi\chi\nu}^2}{4 \pi} \frac{m_\chi^2}{m_{\phi}^4} \,. \label{eq:absorption-xsec}
\end{align}
To assess the potential sensitivity of direct detection to our model, we compare this expression with the projections of Ref.~\cite{Dror:2020czw}. This is justified only within the regime of validity of the EFT, with corrections of order $(m_\chi/m_\phi)^2$ as $m_\phi$ approaches $m_\chi$.

It is evident from \cref{eq:absorption-xsec} that the most interesting parameter space for direct detection is where the mixing is small between $\chi_R'$ and $\nu_R'$ ($\cos\theta_R \simeq 1$) and where the Yukawa coupling in the dark sector is large ($y_{\phi\chi\nu} \simeq 1$). This region of parameter space corresponds to $\vev{\phi'} + M_{\chi\nu} \ll m_\chi$. In subsequent sections we always take $\cos\theta_R = 1$.

%%%%%%%%%%%%%%%%%%%%%%%%%%%%%%

\subsection{Dark Matter Decay}

The dark matter is unstable and can decay into final states featuring neutrinos and photons. There must be an odd number of neutrinos in the final state, since $\chi$ is fermionic. Given the suppression of the decay $\chi\to\nu\gamma$, the dominant decay modes are $\chi\to\nu\gamma\gamma$, and the invisible mode $\chi\to\nu\nu\nu$.

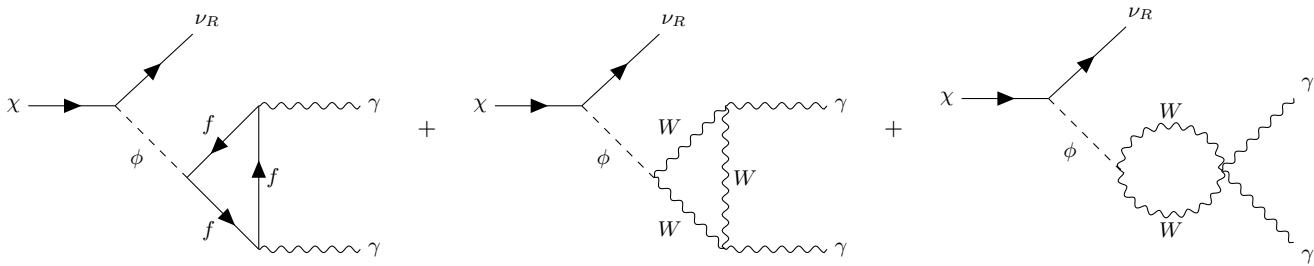
\begin{figure*}
\raisebox{-0.5\height}{
\begin{tikzpicture}
  \begin{feynman}[scale=0.9,transform shape]
    \vertex (a) {\(\chi\)};
    \vertex [right=of a] (b);
    \vertex [above right=of b] (f1) {\(\nu_R\)};
    \vertex [below right=of b] (c);
    \vertex [above right=of c] (d);
    \vertex [right=of d] (f2) {\(\gamma\)};
    \vertex [below right=of c] (e);
    \vertex [right=of e] (f3) {\(\gamma\)};

    \diagram* {
      (a) -- [fermion] (b) -- [fermion] (f1),
      (b) -- [scalar, edge label'=\(\phi\)] (c),
      (c) -- [anti fermion, edge label=\(f\)] (d) -- [anti fermion, edge label = \(f\)] (e) -- [anti fermion, edge label = \(f\)] (c),
      (d) -- [photon] (f2),
      (e) -- [photon] (f3),
      };
  \end{feynman}
\end{tikzpicture}
}
\( \hspace{5pt} + \hspace{5pt} \)
\raisebox{-0.5\height}{
\begin{tikzpicture}
  \begin{feynman}[scale=0.9,transform shape]
    \vertex (a) {\(\chi\)};
    \vertex [right=of a] (b);
    \vertex [above right=of b] (f1) {\(\nu_R\)};
    \vertex [below right=of b] (c);
    \vertex [above right=of c] (d);
    \vertex [right=of d] (f2) {\(\gamma\)};
    \vertex [below right=of c] (e);
    \vertex [right=of e] (f3) {\(\gamma\)};

    \diagram* {
      (a) -- [fermion] (b) -- [fermion] (f1),
      (b) -- [scalar, edge label'=\(\phi\)] (c),
      (c) -- [boson, edge label=\(W\)] (d) -- [boson, edge label = \(W\)] (e) -- [boson, edge label = \(W\)] (c),
      (d) -- [photon] (f2),
      (e) -- [photon] (f3),
      };
  \end{feynman}
\end{tikzpicture}
}
\( \hspace{5pt} + \hspace{5pt} \)
\raisebox{-0.5\height}{
\begin{tikzpicture}
  \begin{feynman}[scale=0.9,transform shape]
    \vertex (a) {\(\chi\)};
    \vertex [right=of a] (b);
    \vertex [above right=of b] (f1) {\(\nu_R\)};
    \vertex [below right=of b] (c);
    \vertex [right=of c] (d);
    \vertex [above right=of d] (f2) {\(\gamma\)};
    \vertex [below right=of d] (f3) {\(\gamma\)};

    \diagram* {
      (a) -- [fermion] (b) -- [fermion] (f1),
      (b) -- [scalar, edge label'=\(\phi\)] (c),
      (c) -- [boson, half left, edge label=\(W\)] (d) -- [boson, half left, edge label = \(W\)] (c),
      (d) -- [photon] (f2),
      (d) -- [photon] (f3),
      };
  \end{feynman}
\end{tikzpicture}
}
\caption{Feynman diagrams for the $\chi\to\nu\gamma\gamma$ decay at one-loop, where $f$ denotes charged SM fermions.}
    \label{fig:FeynmanRadiativeDecay}
\end{figure*}

As discussed in \cref{sec:model}, the softly broken $\mathbb{Z}_2$ symmetry naturally suppresses the otherwise leading visible decay mode $\chi\to\nu\gamma$. In the massless neutrino limit, when $y_{H\nu}$ and hence $\sin\theta_L$ vanish, the only coupling of $\chi$ to the SM is through the scalar portal, and the visible decay modes must proceed via the $\phi-H$ mixing. The $\chi\to\nu\gamma$ decay vanishes in this limit, meaning that the dominant visible decay mode is $\chi\to\nu\gamma\gamma$, which occurs at one-loop via the diagrams shown in \cref{fig:FeynmanRadiativeDecay}. The corresponding decay width is
\begin{equation} \label{eq:chi-vyy}
    \Gamma_{\chi\to\nu\gamma\gamma} = \frac{Y_{\chi\nu}^2\sin^2\theta_\phi}{2048 \pi^5 m_\chi^3}\frac{\alpha^2}{v^2} I_{vis}(m_\phi,m_\chi) \,,
\end{equation}
with
\begin{equation}
    I_{vis}(m_\phi,m_
    \chi) = \left(\frac{11}{6}\right)^2 \int_0^{m_\chi^2} \text{d}s_{12}\left(\frac{m_\chi^2 - s_{12}}{m_\phi^2 - s_{12}}\right)^2 s_{12}^2 \,.
\end{equation}
The factor of $(11/6)^2$ comes from the loop function in the limit where $m_\chi$ is much smaller than the SM fermion and $W$ boson masses. In the regime where $m_\phi \gg m_\chi$, the width can be written purely in terms of the dark matter mass and effective absorption cross-section\footnote{This result differs by a factor of $(11/4)^2$ from the EFT result in Ref.~\cite{Ge:2022ius}, since in that case only electrons run in the loop.}: 
\begin{equation}
    \Gamma_{\chi\to\nu\gamma\gamma} \simeq 1.8 \times 10^{-27} \text{sec}^{-1} \left(\frac{m_\chi}{10\text{ keV}}\right)^5 \left(\frac{\sigma_e}{10^{-50}\text{ cm}^2}\right) \,.
\end{equation}
This provides a good approximation to \cref{eq:chi-vyy}, up to an $\mathcal{O}(1)$ factor, even when $m_\phi \sim m_\chi$. This decay rate must be less than the upper bound imposed by diffuse X-ray measurements~\cite{Essig:2013goa}. 

\begin{figure}
    \includegraphics[width=0.49\textwidth]{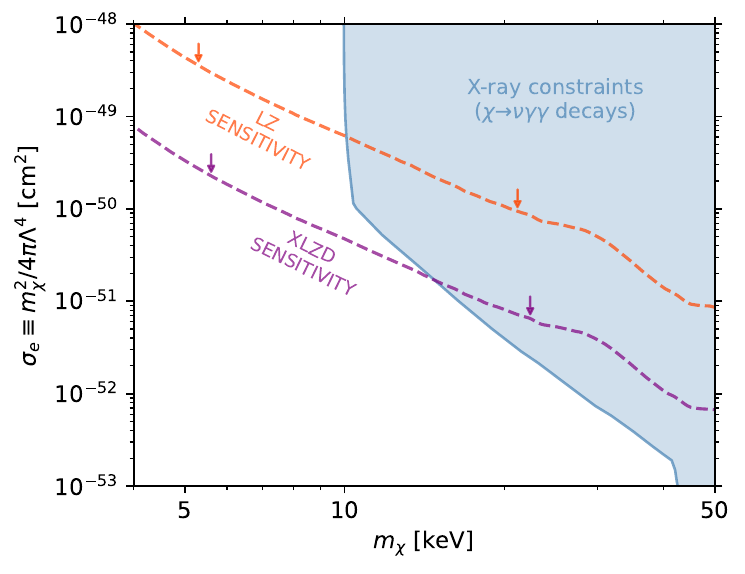}
    \caption{\label{fig:EFT_param_space} Constraints on the dark matter absorption effective cross-section from X-ray searches for $\chi\to\nu\gamma\gamma$ decays, with the excluded region shaded blue. The projected direct detection sensitivities of LZ and XLZD are shown in orange and purple, respectively.}
\end{figure}

\Cref{fig:EFT_param_space} shows the X-ray limits on $\chi\to\nu\gamma\gamma$ decays in terms of the dark matter absorption effective cross-section as a function of $m_\chi$. The sharp changes in the excluded region (shaded blue) near $m_\chi=10,\,40\,$keV arise from the differing sensitivities of the X-ray telescopes used to set the bound. Also shown in the figure is the projected direct detection sensitivity of LZ (assuming 15.3 tonne-years exposure) and the proposed next generation experiment XLZD (assuming 200 tonne-years exposure) from Ref.~\cite{Dror:2020czw}. We will subsequently focus on the sensitivity of XLZD as it offers the best prospects in the foreseeable future. We find that, given the X-ray bounds, XLZD will only provide new sensitivity for dark matter masses $m_\chi \lesssim 15$\,keV. In subsequent sections, we analyse a benchmark dark matter mass of $m_\chi = 10$\,keV.

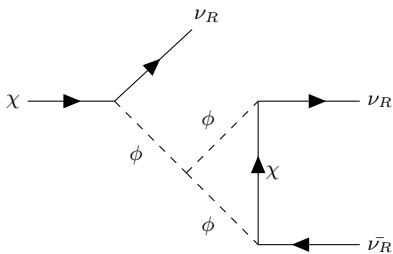
\begin{figure}
    \begin{tikzpicture}
      \begin{feynman}[scale=0.9,transform shape]
        \vertex (a) {\(\chi\)};
        \vertex [right=of a] (b);
        \vertex [above right=of b] (f1) {\(\nu_R\)};
        \vertex [below right=of b] (c);
        \vertex [above right=of c] (d);
        \vertex [right=of d] (f2) {\(\nu_R\)};
        \vertex [below right=of c] (e);
        \vertex [right=of e] (f3) {\(\bar{\nu_R}\)};
    
        \diagram* {
          (a) -- [fermion] (b) -- [fermion] (f1),
          (b) -- [scalar, edge label' = \(\phi\)] (c),
          (c) -- [scalar, edge label = \(\phi\)] (d),
          (c) -- [scalar, edge label' = \(\phi\)] (e),
          (f2) -- [anti fermion] (d) -- [anti fermion, edge label = \(\chi\)] (e) -- [anti fermion] (f3),
          };
      \end{feynman}
    \end{tikzpicture}
    \caption{Feynman diagram for the $\chi\to\nu\nu\nu$ decay at one-loop in the massless neutrino limit ($\sin\theta_L\to0$).}
    \label{fig:FeynmanInvisibleDecay}
\end{figure}

We now turn to the invisible decay, $\chi\to\nu\nu\nu$. This decay can occur at tree level when neutrinos are massive, as can be seen from \cref{fig:absorption-diagram} by replacing the electrons with neutrinos and employing crossing symmetry. The tree-level diagram vanishes in the massless neutrino limit, as do one-loop diagrams featuring $\nu_L$ in the final state. However, the invisible decay into three $\nu_R$ can still occur at one-loop via the $\phi^3$ coupling, as shown in \cref{fig:FeynmanInvisibleDecay}.

The decay width of $\chi\to\nu\nu\nu$ is given by
\begin{equation} \label{eq:chi-vvv}
    \Gamma_{\chi\to\nu\nu\nu} = \frac{Y_{\chi\nu}^6 K^2}{16 m_\chi (4\pi)^7} I_{invis}(m_\phi,m_\chi) \,,
\end{equation}
where, to first order in $\sin\theta_\phi$,
\begin{gather}
    K = \kappa_3 + \kappa_4\vev{\phi'} + 3\delta_2v\sin\theta_\phi \,, \\
    I_{invis}(m_\phi,m_\chi) = \int_0^{m_\chi^2} \text{d}s_{12}\left(\frac{m_\chi^2 - s_{12}}{m_\phi^2 - s_{12}}\right)^2 s_{12} |C_0|^2 \,,
\end{gather}
and $C_0 = C_0(0,0,s_{12},m_\phi^2,m_\chi^2,m_\phi^2)$ is a one-loop scalar integral~\cite{tHooft:1978jhc}. In the regime where $m_\phi \gg m_\chi$, $|C_0|^2 \simeq 1/m_\phi^4$ and the width can be written as
\begin{align}
    \Gamma_{\chi\to\nu\nu\nu} & \simeq 2.5 \times 10^{-21} \text{sec}^{-1} \left(\frac{y_{\phi\chi\nu}\cos\theta_R}{1}\right)^6 \nonumber \\
    & \quad \times \left(\frac{K/m_\phi}{10^{-11}}\right)^2 \left(\frac{m_\chi}{10\text{ keV}}\right)^7 \left(\frac{200 \text{ keV}}{m_\phi}\right)^6.
\end{align}
This differs from \cref{eq:chi-vvv} only by an $\mathcal{O}(1)$ factor, even when $m_\phi \sim m_\chi$. For successful structure formation, $(\Gamma_{\chi\to\nu\nu\nu})^{-1}$ must be greater than 468\,Gyr \cite{FrancoAbellan:2021sxk}. This places an upper bound on the $\phi^3$ coupling of $K \lesssim 10^{-11} m_\phi$. Neglecting possible cancellations between the terms in $K$, this translates into strong upper bounds on the scalar parameters:
\begin{gather}
    |\kappa_3| \lesssim 10^{-6} \text{ eV} \left(\frac{m_\phi}{200\text{ keV}}\right)\label{eq:invis_scalar_bound1} \,,\\
    |\kappa_4\vev{\phi'}| \lesssim 10^{-6} \text{ eV} \left(\frac{m_\phi}{200\text{ keV}}\right)\label{eq:invis_scalar_bound2} \,,\\
    |\delta_2| \lesssim 3\times10^{-11}\left(\frac{10^{-7}}{\sin\theta_\phi}\right)\left(\frac{m_\phi}{200 \text{ keV}}\right) \label{eq:invis_scalar_bound3}.
\end{gather}
Simultaneously satisfying these bounds and obtaining a value of $\sin\theta_\phi$ that is large enough to be of phenomenological interest likely requires values of the scalar potential parameters that are unstable under radiative corrections. Naturalness aside, the above bounds do not restrict the prospects for direct detection. The scalar parameters relevant for direct detection are $m_\phi$, $\sin^2\theta_\phi$, and $\vev{\phi'}$ (which leads to $\chi_R-\nu_R$ mixing). The first two of these are unconstrained by the invisible decay bound. Specifically, $m_\phi^2 \sim \kappa_2$ and $\sin^2\theta_\phi \sim \delta_1^2/m_h^2$ when \cref{eq:invis_scalar_bound1,eq:invis_scalar_bound2,eq:invis_scalar_bound3} are satisfied. Furthermore, the bound on $|\kappa_4\vev{\phi'}|$ imposed by \cref{eq:invis_scalar_bound2} is consistent with the limit $\vev{\phi'} \ll m_\chi$ in which the direct detection prospects are maximised (i.e. $y_{\phi\chi\nu}\simeq1$ and $\cos\theta_R\simeq1$).

%%%%%%%%%%%%%%%%%%%%%%%%%%%%%%

\subsection{Mediator Constraints}

There are also strong constraints on the scalar mediator $\phi$ which, as we shall show, further restrict the parameter space that could be probed through direct detection. The most relevant constraints come from astrophysics. Production of $\phi$ can occur in stars and the core of supernovae, leading to energy loss as $\phi$ or its decay products, $\chi$ and $\nu_R$, escape the stellar medium. We utilise the limits derived in Refs.~\cite{Dev:2020eam, Yamamoto:2023zlu}. Due to the fast decay of $\phi\to\chi\nu_R$ in our model, trapping of $\phi$ within a stellar or supernova core does not occur, however the decay products $\chi$ and $\nu_R$ can be trapped due to the process $\chi N \leftrightarrow \nu_R N$, depending on the value of $y_{\phi\chi\nu}$. We calculate the thermally averaged mean free path for this process using the same stellar/supernova parameters as Refs.~\cite{Dev:2020eam, Yamamoto:2023zlu}.

\begin{figure*}
    \includegraphics[width=0.7\textwidth]{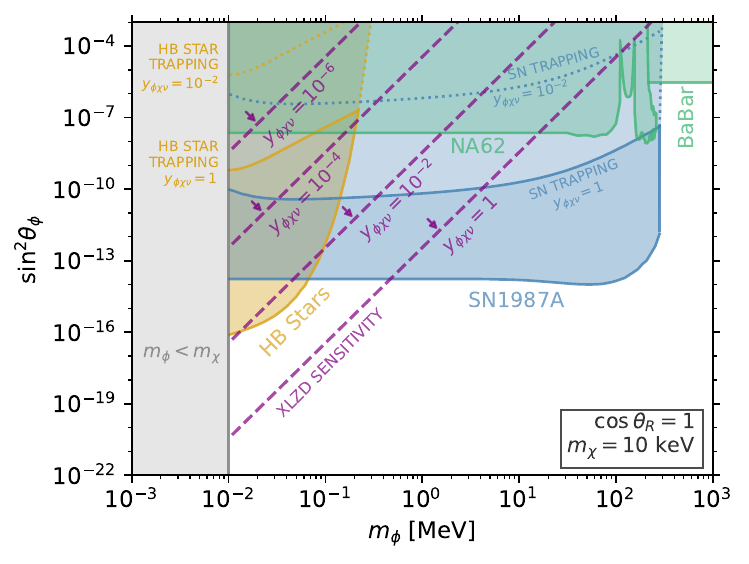}
    \caption{\label{fig:scalar_param_space} Projected direct detection sensitivity of XLZD for $m_\chi=10$\,keV, $\cos\theta_R=1$ and various values of $y_{\phi\chi\nu}$. The yellow and blue shaded regions are excluded by HB star and SN1987A luminosity constraints, respectively. Trapping within the core of these objects depends on $y_{\phi\chi\nu}$ and limits are shown for $y_{\phi\chi\nu}=1$ (solid curves) and $y_{\phi\chi\nu}=10^{-2}$ (dotted curves). See main text for details. Also shown are constraints from meson decays (green), and dark matter stability (grey).}
\end{figure*}

\Cref{fig:scalar_param_space} shows the bounds from stellar and SN1987A luminosity on the $m_\phi - \sin^2\theta_\phi$ parameter space. The upper edges of these regions correspond to where the mean free path of $\chi$ or $\nu_R$ is less than 10\% of the stellar or supernova core radius. This depends on $y_{\phi\chi\nu}$ and we show results for $y_{\phi\chi\nu}=1$ and $y_{\phi\chi\nu}=10^{-2}$. Note that obtaining a precise bound in the trapping regime would require a more detailed treatment. Also shown in \cref{fig:scalar_param_space} are constraints on the scalar mediator from searches for invisible meson decays. The fixed target experiment NA62 constrains the kaon decay $K\to\pi\phi$~\cite{NA62:2021zjw}, while BaBar constrains $B\to K\phi$~\cite{Anchordoqui:2013bfa}. BaBar provides the strongest constraint in the $m_\phi \simeq 300$\,MeV -- 1\,GeV region, where the supernova core is not hot enough to produce $\phi$. The region where $m_\phi < m_\chi$ is excluded by dark matter stability, since here the decay channel $\chi\to\phi\nu$ is available.

For $y_{\phi\chi\nu} \lesssim 0.1$, the combination of SN1987A and NA62 excludes $\sin^2\theta_\phi \gtrsim 10^{-14}$ for $m_\phi \lesssim 300$~MeV, with horizontal branch stars providing stronger limits for $m_\phi \lesssim 70$\,keV. If $y_{\phi\chi\nu} \gtrsim 0.1$, there may be unconstrained windows of parameter space between the NA62 and stellar/SN limits.

Overlaid on \cref{fig:scalar_param_space} is the projected sensitivity of XLZD to $m_\chi = 10$\,keV for a range of $y_{\phi\chi\nu}$ couplings. We see that much of the parameter space accessible to direct detection has already been probed by searches for the mediator. New regions of parameter space are accessible by XLZD, provided $y_{\phi\chi\nu} \gtrsim 10^{-2}$. These results are qualitatively the same across the relevant mass range, 4\,keV $\lesssim m_\chi \lesssim 15$\,keV. As mentioned, corrections of order $(m_\chi/m_\phi)^2$ apply to the projected XLZD sensitivities as $m_\phi$ approaches $m_\chi$ and the EFT begins to break down.

%%%%%%%%%%%%%%%%%%%%%%%%%%%%%%%%%%%%%%%%%%%%%%%%%%%%

\section{\label{sec:production}Dark Matter Production}

We now turn to the production of dark matter in the early universe. To avoid conflicting with bounds on the effective number of relativistic degrees of freedom during BBN and at the time of the CMB~\cite{Boehm:2013jpa}, we require that the dark sector never enters into equilibrium with the SM bath. On the other hand, $\chi$, $\phi$, and $\nu_R$ may achieve thermal equilibrium amongst themselves. This is expected to be the case in the regions of parameter space of interest for direct detection, where the Yukawa coupling in the dark sector is relatively large, $y_{\phi\chi\nu}\gtrsim10^{-2}$ (see \cref{fig:scalar_param_space}). Given some initial population, for example produced via freeze-in, the dark sector reaches equilibrium and subsequent freeze-out in the dark sector sets the final dark matter relic abundance.\footnote{The relic abundance can be produced purely by freeze-in from the SM sector in the parameter space where $y_{\phi\chi\nu}$ is highly suppressed. However, this would not yield a direct detection signal.} This process of freeze-in, thermalisation and freeze-out in the dark sector was first explored in the context of a dark photon mediator in Ref.~\cite{Chu:2011be} and more recently in a scalar portal model in Ref.~\cite{Krnjaic:2017tio}. In the following two sections we discuss in turn thermalisation and freeze-out in detail.

%%%%%%%%%%%%%%%%%%%%%%%%%%%%%%

\subsection{Freeze-in and Thermalisation}\label{sec:production_FI}

The dark sector will be populated through freeze-in from the SM thermal bath, via processes such as $f\bar{f} \to \chi \bar{\nu}_R$, where $f$ denotes SM fermions. This freeze-in production will continue until the SM bath temperature, $T_{SM}$, falls below the electron mass and the process $e^+e^-\to\chi\bar{\nu}_R$ ceases to be efficient. In addition to this irreducible contribution, there may be additional processes that populate the dark sector, for example inflaton decay during reheating.

For the larger values of the scalar mixing that we consider, $\sin^2\theta_\phi \gtrsim 10^{-14}$, the reheat temperature should be well below the electroweak scale. This is because for higher reheat temperatures, electroweak processes, such as $W Z \to W \phi$, can cause the dark sector to enter equilibrium with the SM bath. We use \texttt{micrOMEGAS-v5.3}~\cite{Alguero:2022inz} to numerically calculate the irreducible population of dark sector particles created from (tree-level) freeze-in processes, and estimate that for $\sin^2\theta_\phi = 10^{-14}$, a reheat temperature of $T_R \lesssim 20$~GeV is required. For smaller values of $\sin^2\theta_\phi$, the reheat temperature can be higher, and for $\sin^2\theta_\phi \lesssim 10^{-16}$ the dark sector remains out of equilibrium with the SM regardless of the reheat temperature.

As the dark sector is populated, $\phi, \chi$ and $\nu_R$ interact through the $y_{\phi\chi\nu}$ coupling, which can lead to their thermalisation. Determining whether the dark sector reaches internal equilibrium, and the value of the dark sector temperature $T_D$, requires solving the Boltzmann equations for the momentum distributions generated by freeze-in. However, thermalisation is expected to occur rapidly in the parameter space of interest for direct detection due to the large values of the coupling ($10^{-2} \lesssim y_{\phi\chi\nu} \lesssim 1$). Hence, in the following we simply assume that equilibrium is achieved and consider the temperature in the dark sector as a free parameter. To support this assumption, in \cref{app:dark_EQ} we compare the dark sector interaction rates to the expansion rate and show that for the relevant range of dark sector temperatures and couplings, kinetic and chemical equilibrium are maintained until $T_{SM} \sim m_e/3$ and all freeze-in processes have effectively ceased.

The thermalised dark sector particles contribute to the effective number of relativistic degrees of freedom by an amount that depends on the temperature ratio $T_D/T_{SM}$. The limits on $\Delta N_{eff}$ from both BBN and CMB~\cite{Yeh:2022heq} translate to an upper bound on the dark sector temperature of $T_D \lesssim 0.4 \times T_{SM}$.

%%%%%%%%%%%%%%%%%%%%%%%%%%%%%%

\subsection{Freeze-out of Dark Matter}\label{sec:production_FO}

\begin{figure}
    \includegraphics[width=0.49\textwidth]{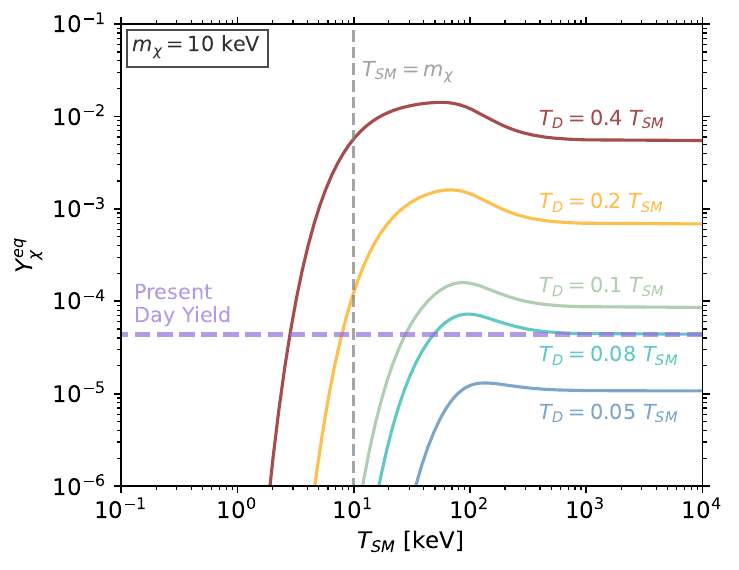}
    \caption{\label{fig:Y_eq_TD} Yield of $\chi$ while in equilibrium with a dark sector of temperature $T_D$, as a function of the SM bath temperature, $T_{SM}$. The various curves are for different dark sector temperatures, with warmer (colder) colours indicating a warmer (colder) dark sector. The horizontal purple dashed line indicates the observed present day dark matter yield. If $\chi$ is to freeze-out with the correct abundance, the dark sector must be sufficiently warm. Results are shown for $m_\chi = 10$\,keV.}
\end{figure}

Next, we calculate the relic abundance of dark matter produced via freeze-out within the dark sector. The equilibrium number density of $\chi$ at a dark sector temperature $T_D$ is given by
\begin{equation}\label{eq:n_eq}
    n_\chi^{eq}\left(T_D\right) = \frac{g_\chi m_\chi^2 T_D}{2 \pi^2} K_2\left(m_\chi/T_D\right) \,,
\end{equation}
where $g_\chi = 4$ and $K_2$ is a modified Bessel function of the second kind. We divide the number density by the SM entropy density to obtain the yield, or comoving number density\footnote{After the freeze-in processes have ended, the entropy density of the SM and dark sectors are (approximately) separately conserved.}:
\begin{equation}\label{eq:Yield}
    Y^{eq}_\chi\left(T_D,T_{SM}\right) \equiv \frac{n^{eq}_\chi(T_D)}{s(T_{SM})} \,.
\end{equation}
\Cref{fig:Y_eq_TD} shows the equilibrium yield of $\chi$ for various ratios of $T_D/T_{SM}$ as a function of $T_{SM}$. The observed present day yield of $\chi$, shown as a horizontal purple dashed line, must be obtained via freeze-out from these equilibrium curves. Hence, to achieve the correct relic abundance requires $0.08 \lesssim T_D/T_{SM} \lesssim 0.4$ at the onset of freeze-out, with the upper limit coming from the bound on $\Delta N_{eff}$. 

We now solve the Boltzmann equation across the viable range of $T_D/T_{SM}$, to find the range of $y_{\phi\chi\nu}$ that gives the observed present day dark matter abundance. The Boltzmann equation for the dark matter number density, $n_\chi(t)$, is
\begin{equation}
    \dot{n}_\chi + 3H n_\chi = -\left<\sigma v\right>_{T_D}\left(n_\chi^2 - (n_\chi^{eq})^2\right) \,,
\end{equation}
where $\left<\sigma v\right>_{T_D}$ is the thermally averaged cross-section times velocity evaluated at temperature $T_D(t)$, and we have assumed Maxwell-Boltzmann statistics. At the time of freeze-out, the relevant number-changing processes are $\chi\chi\to\nu_R\nu_R$, $\bar{\chi}\bar{\chi}\to\bar{\nu}_R\bar{\nu}_R$, and $\chi\bar{\chi}\to\nu_R\bar{\nu}_R$. We use \texttt{micrOMEGAS-v5.3}~\cite{Alguero:2022inz} to numerically calculate $\left<\sigma v\right>$ as a function of temperature. When the $\Delta N_{eff}$ bound is satisfied, the Hubble parameter $H(t)$ is largely determined by the SM degrees of freedom. In addition, the effective number of entropic degrees of freedom is approximately constant during freeze-out, $\mathrm{d}g_{*S}/\mathrm{d}T_{SM} \simeq 0$. Hence, expressed in terms of the yield, $Y_{\chi} \equiv n_\chi/s(T_{SM})$, the Boltzmann equation becomes
\begin{equation}
    \frac{dY_\chi}{dT_{SM}} = \frac{s(T_{SM})}{H(T_{SM})T_{SM}}\left<\sigma v\right>_{T_D}\left(Y_\chi^2 - \left(Y_\chi^{eq}\right)^2\right) \,.
\end{equation}

\begin{figure}
    \includegraphics[width=0.49\textwidth]{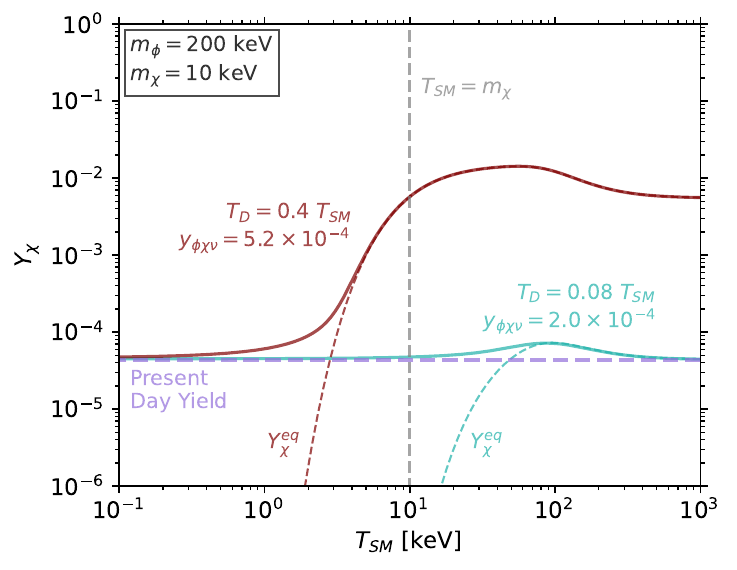}
    \caption{\label{fig:Y_sol} Yield of dark matter $\chi$ when freezing out from a dark sector of temperature $T_D$, as a function of the SM bath temperature, $T_{SM}$. The horizontal purple dashed line indicates the observed present day dark matter yield. The Yukawa coupling $y_{\phi\chi\nu}$ is set for each curve by requiring freeze-out at the correct present day yield. Results are shown for $m_\phi = 200$\,keV, $\cos\theta_R = 1$, and $m_\chi=10$\,keV. Results for other values are discussed in the text.}
\end{figure}

\Cref{fig:Y_sol} shows the freeze out of $\chi$ within the dark sector at the extremal allowed values of $T_D$. The red curve shows the hottest dark sector, $T_D = 0.4 \times T_{SM}$, while the blue curve shows the coldest dark sector, $T_D = 0.08 \times T_{SM}$. In each case the Yukawa coupling $y_{\phi\chi\nu}$ has been chosen to obtain the observed dark matter abundance. For the hotter dark sector, $y_{\phi\chi\nu}$ must be larger so that equilibrium is prolonged and there is sufficient annihilation of dark matter into neutrinos. The solutions reveal that for $\chi$ to account for the present day yield, $y_{\phi\chi\nu}$ must be in the range $2\times10^{-4}$ to $5\times10^{-4}$ for $m_\phi = 200$\,keV. This range varies linearly with $m_\phi$, since $\left<\sigma v\right> \propto \left(y_{\phi\chi\nu}/m_\phi\right)^4$. Furthermore, the required $y_{\phi\chi\nu}$ varies negligibly across the range of $m_\chi$ of interest for direct detection (4\,keV $\lesssim m_\chi \lesssim 15$\,keV). Note that the results are independent of $\sin^2\theta_\phi$, since all interactions with the SM during freeze-out are Boltzmann suppressed.

Finally, we consider the potential of direct detection to probe the values of $y_{\phi\chi\nu}$ that yield the correct dark matter abundance. From \cref{fig:scalar_param_space}, we see that for $m_\phi = 200$\,keV and $2\times10^{-4} < y_{\phi\chi\nu} < 5\times10^{-4}$, the sensitivity of XLZD to $\sin^2\theta_\phi$ lies well within the parameter space already excluded by SN1987A. Furthermore, given that the relevant range of $y_{\phi\chi\nu}$ depends linearly on $m_\phi$, as discussed above, this conclusion holds for the full range of $m_\phi$ in \cref{fig:scalar_param_space}. Hence, after imposing the correct relic abundance, there is no unconstrained parameter space that can be probed via dark matter absorption at XLZD.

%%%%%%%%%%%%%%%%%%%%%%%%%%%%%%%%%%%%%%%%%%%%%%%%%%%%

\section{\label{sec:conclusion}Conclusion}

Fermionic dark matter absorption is an interesting direct detection signal that has recently been studied using effective field theory. Motivated by this, we have constructed a UV complete model of the scalar operator associated with this signature. Our model can achieve the correct cosmological relic abundance via freeze-out within the dark sector. We find that visible dark matter decays, in conjunction with stellar and supernova constraints on the scalar mediator, provide strong constraints. The parameter space that is consistent with these bounds and could also be probed at a next-generation direct detection experiment such as XLZD under-predicts the relic density. This case study demonstrates that the absorption signal cannot be observed in upcoming direct detection experiments for a minimal, realistic model of scalar-mediated dark matter. It would therefore be interesting to construct other, non-minimal models which have this signal as well as a consistent cosmological history yielding the correct relic abundance.

%%%%%%%%%%%%%%%%%%%%%%%%%%%%%%%%%%%%%%%%%%%%%%%%%%%%

\begin{acknowledgments}
This work was supported in part by the Australian Research Council through the ARC Centre of Excellence for Dark Matter Particle Physics, CE200100008 and the Australian Government Research Training Program Scholarship initiative. P.C. is supported by the Australian Research Council Discovery Early Career Researcher Award DE210100446.
\end{acknowledgments}

%%%%%%%%%%%%%%%%%%%%%%%%%%%%%%%%%%%%%%%%%%%%%%%%%%%%

\appendix

\section{\label{app:model_details} Diagonalisation of the Model}

This appendix provides details on the diagonalisation of the scalar and fermion sectors, as well as the full expressions for their interaction terms in the mass basis.

%%%%%%%%%%%%%%%%%%%%%%%%%%%%%%

\subsection{Scalar Sector Diagonalisation}

The full scalar potential is given by
\begin{align}
    V(H,\phi') = & -M_H^2 H^\dag H + \lambda_H(H^\dag H)^2 \notag \\
    & + \kappa_{1} \phi' + \frac{\kappa_2}{2}\phi'^2 + \frac{\kappa_3}{3!}\phi'^3 + \frac{\kappa_4}{4!}\phi'^4 \notag \\
    & + \delta_1 H^\dag H \phi' + \frac{\delta_2}{2}H^\dag H \phi'^2 \,.
\end{align}
We expand $H$ around its VEV with the parameterisation
\begin{equation}
    H \equiv \frac{1}{\sqrt{2}} \begin{pmatrix}0 \\ v + h'\end{pmatrix} \,,
\end{equation}
where $h'$ is a real scalar field. We also re-express $\phi'$ as an expansion around its (possibly zero) VEV: $\phi' \to \phi' + \vev{\phi'}$. Minimising the potential gives
\begin{equation}
    v = \sqrt{\frac{2M_H^2 - 2\delta_1\vev{\phi'} - \delta_2 \vev{\phi'}^2}{2\lambda_H}} \,,
\end{equation}
and the VEV of $\phi'$ satisfies the cubic equation
\begin{equation}\label{eq:phi_cubic}
    \kappa_4\vev{\phi'}^3 + 3\kappa_3\vev{\phi'}^2 + \left(6\kappa_2 + 3\delta_2 v^2\right)\vev{\phi'} + \left(6\kappa_1 + 3\delta_1 v^2\right) = 0 \,.
\end{equation}

The parameterisation of the scalar fields about their VEVs results in the mass terms
\begin{equation}
    V_{mass} = \frac{1}{2}\begin{pmatrix}h' & \phi'\end{pmatrix}
    \begin{pmatrix}\mu_{h'}^2 & \mu_{h' \phi'}^2/2 \\ \mu_{h' \phi'}^2/2 & \mu_{\phi'}^2\end{pmatrix}
    \begin{pmatrix}h' \\ \phi'\end{pmatrix} \,,
\end{equation}
where
\begin{align}
    \mu_{h'}^2 & = 2\lambda_H v^2 \,, \\
    \mu_{\phi'}^2 & = \frac{\delta_2 v^2}{2} + \kappa_2 + \kappa_3 \vev{\phi'} + \frac{\kappa_4}{2} \vev{\phi'}^2 \,, \\
    \mu_{h' \phi'}^2 & = 2v(\delta_1 +\delta_2 \vev{\phi'}) \,.
\end{align}
Diagonalising the mass matrix above yields the mass eigenvalues
\begin{equation}
    m_{h,\phi}^2 = \left(\frac{\mu_{h'}^2 + \mu_{\phi'}^2}{2}\right) \pm \left(\frac{\mu_{h'}^2 - \mu_{\phi'}^2}{2}\right)\sqrt{1 + \left(\frac{\mu_{h' \phi'}^2}{\mu_{h'}^2 - \mu_{\phi'}^2}\right)^2} \,,
\end{equation}
where $h$ and $\phi$ (unprimed) denote the heavier and lighter mass eigenstates respectively. The mixing can be expressed in terms of an angle $\theta_\phi$, such that
\begin{equation}
    \begin{pmatrix}h \\ \phi\end{pmatrix} =
    \begin{pmatrix}\cos\theta_\phi & -\sin\theta_\phi \\ \sin\theta_\phi & \cos\theta_\phi\end{pmatrix}
    \begin{pmatrix}h' \\ \phi'\end{pmatrix} \,,
\end{equation}
with
\begin{equation}\label{eq:tanthetaS}
    \tan 2\theta_\phi = \frac{-\mu_{h' \phi'}^2}{\mu_{h'}^2 - \mu_{\phi'}^2} \,.
\end{equation}
In the limit of small mixing ($\theta_\phi^2 \ll 1$),
\begin{equation}
    \sin^2\theta_\phi \simeq \frac{v^2 (\delta_1+\delta_2\vev{\phi'})^2}{(m_h^2-m_\phi^2)^2} \,.
\end{equation}

%%%%%%%%%%%%%%%%%%%%%%%%%%%%%%

\subsection{Fermion Sector Diagonalisation}

With $\phi'$ and $H$ expanded about their VEVs, the mass terms for $\chi'$ and $\nu'$ are given by
\begin{equation}
    \mathcal{L} \supset -
    \begin{pmatrix}
        \bar{\nu}'_L & \bar{\chi}'_L 
    \end{pmatrix}
    \mathcal{M}
    \begin{pmatrix}
        \nu'_R \\ \chi'_R
    \end{pmatrix} + h.c. \,,
\end{equation}
where
\begin{equation}\label{eq:MMatrix3x3}
    \mathcal{M} \equiv \begin{pmatrix}
        M_\nu & 0 \\
        y_{\phi \chi \nu} \vev{\phi'} + M_{\chi\nu} & M_\chi
    \end{pmatrix} \,,
\end{equation}
and $M_\nu = y_{H\nu}v/\sqrt{2}$. Note that without loss off generality, we can take the three non-zero elements of $\mathcal{M}$ to be real and positive by re-phasing of the fermion fields.

The fermion mass matrix can be diagonalised via a bi-unitary transformation:
\begin{equation}
    \begin{pmatrix}
        m_\nu & 0 \\
        0 & m_\chi
    \end{pmatrix}
    =
    U_L \mathcal{M} U_R^\dag \,,
\end{equation}
where $U_L$ and $U_R$ are $2\times 2$ unitary matrices. Under this transformation, the mass basis fields (denoted without primes) are given by
\begin{align}
    \begin{pmatrix}
        \nu_{L} \\ \chi_{L}
    \end{pmatrix}
    = U_{L} \begin{pmatrix}
        \nu'_{L} \\ \chi'_{L}
    \end{pmatrix} \,,
    &&
    \begin{pmatrix}
        \nu_{R} \\ \chi_{R}
    \end{pmatrix}
    = U_{R} \begin{pmatrix}
        \nu'_{R} \\ \chi'_{R}
    \end{pmatrix} \,.
\end{align}
The masses $m_\nu$ and $m_\chi$ are the singular values of $\mathcal{M}$:
\begin{align}
    m_\nu^2 & = \frac{1}{2}\left[a - \sqrt{a^2 - 4M_\nu^2M_\chi^2}\right] \,, \\
    m_\chi^2 & = \frac{1}{2}\left[a + \sqrt{a^2 - 4M_\nu^2M_\chi^2}\right] \,,
\end{align}
with
\begin{equation}
    a = M_\nu^2 + M_\chi^2 + (y_{\phi\chi\nu}\vev{\phi'}+M_{\chi\nu})^2 \,.
\end{equation}

Since $\mathcal{M}$ has real entries, we can parameterise $U_{L,R}$ in terms of real rotations,
\begin{equation}
    U_{L,R} = \begin{pmatrix}
        \cos\theta_{L,R} & -\sin\theta_{L,R} \\
        \sin\theta_{L,R} & \cos\theta_{L,R}
    \end{pmatrix} \,,
\end{equation}
and the mixing angles can be written in terms of the Lagrangian parameters as
\begin{align}
    \tan2\theta_L &=\frac{2 M_\nu (y_{\phi\chi\nu} \vev{\phi'}+M_{\chi\nu})}{M_\chi^2 - M_\nu^2 + (y_{\phi\chi\nu}\vev{\phi'}+M_{\chi\nu})^2} \,, \\
    \tan2\theta_R &= \frac{2 M_\chi (y_{\phi\chi\nu} \vev{\phi'}+M_{\chi\nu})}{M_\chi^2 - M_\nu^2 - (y_{\phi\chi\nu}\vev{\phi'}+M_{\chi\nu})^2} \,.
\end{align}

In the massless neutrino limit, $\sin\theta_L$ vanishes. The region of parameter space where $\cos\theta_R \simeq 1$ offers the best prospects for direct detection. In this region $y_{\phi\chi\nu}\vev{\phi'} + M_{\chi\nu} \ll M_\chi$, which may require fine-tuning of the scalar potential if $y_{\phi\chi\nu} \sim 1$.

%%%%%%%%%%%%%%%%%%%%%%%%%%%%%%

\subsection{Interaction Terms in the Mass Basis}

Upon transforming to the mass basis, we obtain the following tree-level couplings between $\chi$, $\phi$ and $\nu_{L,R}$:
\begin{equation}
    \mathcal{L} \supset -\phi
    \begin{pmatrix}
        \bar{\nu_L} & \bar{\chi_L} 
    \end{pmatrix}
    \begin{pmatrix}
        Y_{\nu\nu} & Y_{\chi\nu_L} \\
        Y_{\chi\nu_R} & Y_{\chi\chi}
    \end{pmatrix}
    \begin{pmatrix}
        \nu_R \\ \chi_R
    \end{pmatrix} + h.c. \,,
\end{equation}
with,
\begin{align}
    Y_{\nu\nu} &= - y_{\phi\chi\nu}\cos\theta_\phi \sin\theta_L\cos\theta_R \nonumber\\ 
    & \quad\quad\quad\quad + \frac{y_{H\nu}}{\sqrt{2}}\sin\theta_\phi \cos\theta_L\cos\theta_R \,, \\
    Y_{\chi\nu_L} &= - y_{\phi\chi\nu}\cos\theta_\phi \sin\theta_L\sin\theta_R \nonumber\\ 
    & \quad\quad\quad\quad + \frac{y_{H\nu}}{\sqrt{2}}\sin\theta_\phi \cos\theta_L\sin\theta_R \,, \\
    Y_{\chi\nu_R} &= y_{\phi\chi\nu}\cos\theta_\phi \cos\theta_L\cos\theta_R \nonumber\\ 
    & \quad\quad\quad\quad + \frac{y_{H\nu}}{\sqrt{2}}\sin\theta_\phi \sin\theta_L\cos\theta_R \,, \\
    Y_{\chi\chi} &= y_{\phi\chi\nu}\cos\theta_\phi \cos\theta_L\sin\theta_R \nonumber\\ 
    & \quad\quad\quad\quad + \frac{y_{H\nu}}{\sqrt{2}}\sin\theta_\phi \sin\theta_L\sin\theta_R \,.
\end{align} 
In the massless neutrino limit, where $y_{H\nu}$ and $\sin\theta_L$ vanish, the couplings simplify to
\begin{gather}
    Y_{\nu\nu} = 0 \,, \\
    Y_{\chi\nu_L} = 0 \,, \\
    Y_{\chi\nu_R} = y_{\phi\chi\nu}\cos\theta_\phi \cos\theta_R \,, \\
    Y_{\chi\chi} = y_{\phi\chi\nu}\cos\theta_\phi \sin\theta_R \,,
\end{gather}
which is the result stated in \cref{sec:model}.

%%%%%%%%%%%%%%%%%%%%%%%%%%%%%%%%%%%%%%%%%%%%%%%%%%%%

\section{\label{app:dark_EQ} Dark Sector Equilibrium}

If the dark sector remains in chemical and kinetic equilibrium after all freeze-in processes have effectively ceased, the freeze-in and freeze-out eras of dark matter production can be treated separately, as in \cref{sec:production}. In this appendix, we show that chemical and kinetic equilibrium are indeed maintained post-freeze-in for the range of dark sector temperatures and couplings that allow for the freeze-out of $\chi$ with the correct relic abundance.

\begin{figure}
    \includegraphics[width=0.49\textwidth]{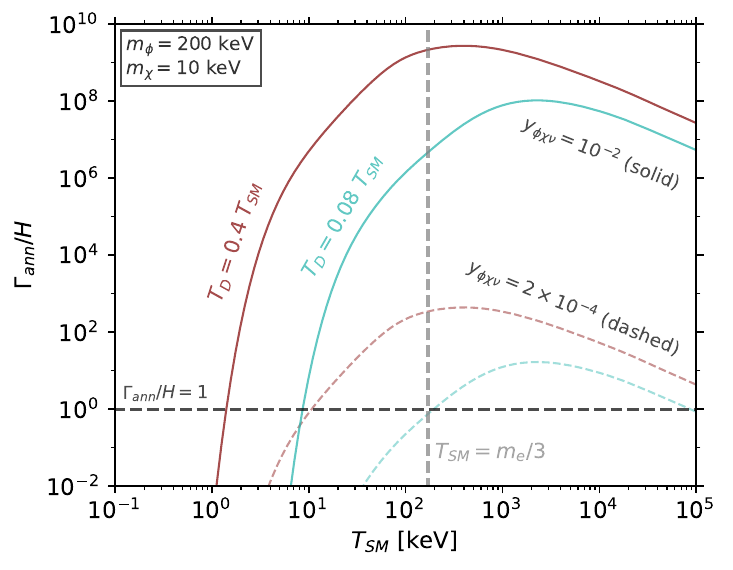}
    \caption{\label{fig:DS_eq} The dark matter annihilation rate compared to the Hubble rate in the early universe. Red and green curves correspond to $T_D/T_{SM} = 0.4$ and $0.08$, respectively. Solid curves show results for $y_{\phi\chi\nu} = 10^{-2}$ and dashed curves for $y_{\phi\chi\nu} = 2\times 10^{-4}$. For all curves, $m_\chi = 10$\,keV and $m_\phi = 200$\,keV. The vertical line at $T_{SM} = m_e/3$ indicates the temperature below which all freeze-in processes have effectively concluded.}
\end{figure}

Chemical equilibrium is maintained between $\chi$ and $\nu_R$ when the annihilation rate of dark matter to sterile neutrinos is greater than the Hubble parameter. We calculate the ratio of these two quantities as a function of $T_D$ and $T_{SM}$:
\begin{equation}\label{eq:DS_rate}
    \frac{\Gamma_{\mathrm{ann}}}{H} = \frac{n_\chi^{eq}(T_D) \left<\sigma v \right>_{T_D}}{H(T_{SM})} \,,
\end{equation}
where $n_\chi^{eq}(T_D)$ is the equilibrium number density of $\chi$ and $\left<\sigma v\right>_{T_D}$ is the thermally averaged dark matter annihilation cross-section times velocity at temperature $T_D$. As in \cref{sec:production}, we use \texttt{micrOMEGAS-v5.3}~\cite{Alguero:2022inz} to numerically calculate $\left<\sigma v\right>$, including contributions from $\chi\chi\to\nu_R\nu_R$, $\bar{\chi}\bar{\chi}\to\bar{\nu}_R\bar{\nu}_R$, and $\chi\bar{\chi}\to\nu_R\bar{\nu}_R $. We work in the limit of massless neutrinos and take $\cos\theta_R = 1$. The $\Delta N_{eff}$ bound discussed in \cref{sec:production} enforces $T_D < 0.4 \times T_{SM}$, such that the Hubble rate is largely determined by the SM degrees of freedom. Kinetic equilibrium is maintained whenever chemical equilibrium holds and persists to lower temperatures, since the annihilation rate is suppressed by a factor of $n_\chi^{eq}$ compared with the scattering rate $\chi\nu_R\leftrightarrow\chi\nu_R$.

\Cref{fig:DS_eq} shows $\Gamma_{\mathrm{ann}}/H$ as a function of $T_{SM}$ for two dark sector couplings, $y_{\phi\chi\nu} = 10^{-2}$ and $2\times 10^{-4}$, and two dark sector temperatures, $T_D/T_{SM} = 0.4$ and 0.08. These temperatures correspond to the upper and lower bounds discussed in \cref{sec:production_FO}, between which successful freeze-out of $\chi$ with the correct relic abundance can occur. At SM bath temperatures below $T_{SM} \sim m_e/3$, shown as a vertical grey dashed line, freeze-in processes from the SM bath are Boltzmann suppressed. We see that for $0.08 \lesssim T_D/T_{SM} \lesssim 0.4$, dark sector equilibrium is maintained both during and beyond the end of the freeze-in era, provided $y_{\phi\chi\nu} \gtrsim 2\times 10^{-4}$. This supports the assumption made in \cref{sec:production} that the dark sector enters equilibrium as $\phi$, $\chi$ and $\nu_R$ are produced via portal interactions, and that the freeze-in and freeze-out eras of dark matter production can be treated separately.

%%%%%%%%%%%%%%%%%%%%%%%%%%%%%%%%%%%%%%%%%%%%%%%%%%%%

\bibliography{DMA}

\end{document}